\documentclass[nofootinbib,showpacs,aps,twocolumn,preprintnumbers,letterpaper]{revtex4-2}
\usepackage{amsmath,amssymb}
\usepackage{epsfig}
\usepackage{graphicx}
\usepackage{amsmath}
\usepackage{slashed}
\usepackage{amsfonts}
\usepackage{epstopdf}

\usepackage{graphicx,subfigure}
\usepackage{amsmath,amssymb}
\usepackage{hhline}
\usepackage[pdftex, pdfborder= 0 0 0, citecolor=blue, urlcolor=blue, linkcolor=blue, colorlinks=true, bookmarksopen=true]{hyperref}

\newcommand{\be}{\begin{eqnarray}}
\newcommand{\ee}{\end{eqnarray}}

\def\slashchar#1{\setbox0=\hbox{$#1$}           
   \dimen0=\wd0                                 
   \setbox1=\hbox{/} \dimen1=\wd1               
  \ifdim\dimen0>\dimen1                        
 \rlap{\hbox to \dimen0{\hfil/\hfil}}      
  #1                                        
 \else                                        
    \rlap{\hbox to \dimen1{\hfil$#1$\hfil}}   
    /                                         
 \fi}                                         %

\begin{document}

\title{Chiral Symmetry Breaking and Confinement from an Interacting Ensemble of Instanton-dyons in Two-flavor Massless QCD}

\author{ Dallas DeMartini  and Edward  Shuryak }

\affiliation{Center for Nuclear Theory, Department of Physics and Astronomy, Stony Brook University,
Stony Brook, NY 11794-3800, USA}

\begin{abstract}
	In this work we present the results from numerical simulations of an interacting ensemble of instanton-dyons in the $SU(3)$ gauge group with $N_f=2$ flavors of massless quarks. Dynamical quarks are included via the effective interactions induced by the fermionic determinant evaluated in the subspace of topological zero modes. The eigenvalue spectrum of the Dirac operator is studied at different volumes to extract the chiral condensate and eigenvalue gap, with both observables providing consistent values of the chiral transition temperature $T_c$. We find that a sufficient density of dyons is responsible for generating the confining potential and breaking the chiral symmetry, both of which are compatible with second-order transitions.   
\end{abstract}
\maketitle

\section{Introduction}

\subsection{Instantons-dyons at finite temperature}

Quantum Chromodynamics (QCD) possesses approximate symmetries whose breaking/restoration correspond to certain phase transitions. At finite temperature and zero chemical potentials, QCD has two crossovers: deconfinement and chiral symmetry restoration. In the infinite quark mass (pure gauge) limit, the theory has an exact $\mathbb{Z}_3$ symmetry, for which the average Polyakov loop $\langle P \rangle$ is an order parameter. (The Polyakov loop $\hat{P}$ is a matrix and we define the average Polyakov loop as the scalar quantity $\langle P \rangle = \frac{1}{3} \langle  Tr[\hat{P}(\vec{x})] \rangle$.) At finite quark masses, it is no longer a strict order parameter, but still has a rapid change and a psuedocritical temperature that can be identified from its inflection point. 

The average Polyakov loop is associated with the deconfinement of quark degrees of freedom by its relation to the heavy quark free energy \cite{Kaczmarek:2002mc}
\begin{equation}
	\langle P \rangle = \exp (-F_Q/T).
\end{equation}
Of course, at large $T$ the quarks are asymptotically free with trivial Polyakov loop $\langle P \rangle \rightarrow 1$. The question is then: which nonperturbative interactions drive the theory to the confining Polyakov loop $\langle P \rangle = 0$ at low temperatures?

The instanton in $SU(N)$ gauge theories is the minimum of the action with vanishing fields $A_{\mu}^a$ at space-time infinity. 
 At finite temperature, with  a nonzero VEV  of the Polyakov loop,  one component along the Euclidean time direction is nonzero $\langle A_4 \rangle \ne 0$. Looking for solutions of Yang-Mills equations with such modified conditions at infinity, it was found that the instanton dissolves into $N_c$ constituent dyons (also known as instanton-monopoles) connected by Dirac strings \cite{Lee:1998bb,Kraan:1998sn}. Unlike the instantons, the dyons 
 interact directly with the holonomy. It was suggested then that the dyons can generate a confining potential which overcomes the perturbative interactions of thermal gluons. For a review, see e.g. Ref. \cite{Diakonov:2009jq}.

Analytic descriptions of how the dyons generate confine are possible in particular supersymmetric models \cite{Poppitz:2011wy,Poppitz:2012sw}, which can be achieved with a dilute gas of dyons due to a cancellation of the deconfining potential. In the standard Yang-Mills theories, the deconfining Gross-Pisarski-Yaffe potential \cite{Gross:1980br} means that a dense, strongly-coupled ensemble is required to confine the theory. In these cases, the dyons have been shown to generate confinement via numerical simulations in both the pure $SU(2)$ \cite{Larsen:2015vaa,Lopez-Ruiz:2016bjl} and, more recently, $SU(3)$ \cite{DeMartini:2021dfi} cases.

One can also take the 'inverse' approach, identifying dyons in lattice configurations using the Dirac operator and gradient flow methods to reveal the dyons \cite{Gattringer:2002wh,Bornyakov:2015xao,Larsen:2019sdi} from underneath the quantum fluctuations of the gluon field.

\subsection{Instanton-dyons and fermions}

If two light quark flavors are massless, QCD has an exact $SU(2)_L \times SU(2)_R$ chiral flavor symmetry.  
Below  some $T_c$ the axial part of this symmetry 
is spontaneously broken. The chiral quark condensate $\langle \bar{q}q \rangle$ serves as the order parameter of the transition. 

The interactions between instantons and quarks have been studied since the 1980's, starting with the instanton-induced 't Hooft Lagrangian. This interaction explicitly breaks the $U_A(1)$ symmetry via the topological quark zero modes. Later, the Instanton Liquid Model (ILM) \cite{Shuryak:1982dp} showed that the breaking of chiral symmetry is related to the collectivization of said zero modes into the so-called zero-mode zone (ZMZ). For a review, see e.g. Ref. \cite{Schafer:1996wv}. 

Following the discovery of the instanton dyons, it was shown that the quark zero mode of the instanton localizes to a single constituent dyon. Which one depends on the quark periodicity condition \cite{GarciaPerez:1999ux}. These zero modes, like the dyons themselves, have explicit dependence on the holonomy. In the case of (physical) antiperiodic quarks, all zero modes are localized to the $L$ dyons as will be discussed in the next sections. 

Previous studies of ensembles of dyons have analyzed the Dirac eigenvalue spectrum in this zero-mode zone and confirmed their role in chiral symmetry breaking in the $SU(2)$ gauge group via mean-field methods \cite{Liu:2015jsa} and numerical studies \cite{Larsen:2015tso,LopezRuiz:2019hvh}. Both techniques have also been employed to study the phase transitions in theories with modified quark periodicities \cite{Larsen:2016fvs,Liu:2016yij}. The goal of this work is then to extend such numerical studies, in an effort to approach physical QCD, to the $SU(3)$ gauge group with $N_f=2$ flavors of massless, dynamical quarks, looking at both the confinement and chiral symmetry transitions. 

More recently dyons have been identified on the lattice via their quark zero modes \cite{Larsen:2019sdi}. It has been shown that the form of the quark zero modes are remarkably insensitive to the many perturbative gluons in which they are submerged. From studies such as this, dyon densities and correlation functions can be calculated, serving as useful constraints on models of the dyon interactions such as ours. 

The structure of this paper is as follows: In Section \ref{sec_partition} the physics of the dyon ensemble, their interactions, and the holonomy is described. Section \ref{sec_quark} focuses on the quark zero modes and the interactions induced by them. Technical details of the simulations are discussed in Section \ref{sec_tech}. Finally, Sections \ref{sec_poly} and \ref{sec_chiral} lay out the results relevant to the deconfinement and chiral phase transitions, respectively.  

\section{Interacting Dyon Ensemble} \label{sec_partition}
\subsection{Holonomy and the dyon partition function} 

The instanton-dyons are obtained by generalizing the instanton solution to nonzero holonomy (nontrivial Polyakov loop) \cite{Kraan:1998sn,Lee:1998bb}. In $SU(3)$, the instanton is decomposed into three dyon species, the $M_1$ and $M_2$ dyons corresponding to the maximally diagonal subgroup and the 4-time dependent $L$ dyon, as well as corresponding antidyons. The holonomies are the differences in the phases $\mu_1$, $\mu_2$, $\mu_3$ of the eigenvalues of the gauge field component $A_4$ at spatial infinity. The holonomies are defined as $\nu_i = \mu_{i+1} - \mu_i$ with $\sum \nu_i = 1$. Demanding that $\langle P \rangle$ be real reduces the individual dyon holonomies to depend on a single parameter $\nu$.

The dyon actions and core sizes are directly related to their individual holonomies. The actions of each dyon, in terms of the instanton action $S_0$, are
\begin{equation}
S_{M1} = S_{M2} = S_0, \nu \,\,\,\,\, S_L = S_0 (1-2\nu) = S_0 \bar{\nu}.
\end{equation}
The sizes of the dyon cores scale as $1/\nu_i$.

We work in dimensionless units with all lengths in units of $1/T$, and define the following dimensionless quantities: the volume $\tilde{V}_3 = (LT)^3$, free energy $\tilde{F}= F/T$, free energy density $f= \tilde{F}/\tilde{V}_3$, and dyon densities $n_i = N_i/\tilde{V}_3$. The single holonomy parameter is defined on the interval $\nu \in [0,1/2]$ and is related to the Polyakov loop via
\begin{equation}
\langle P \rangle = \frac{1}{3} + \frac{2}{3} \cos (2 \pi \nu).
\end{equation} 
The main scale of interest, the instanton action $S_0$, is related to the temperature by 
\begin{equation}
S_0 = \frac{8 \pi^2}{g^2} = \big( \frac{11}{3}N_c - \frac{2}{3} N_f \big) \ln (T/\Lambda),
\end{equation} 
where $N_c = 3$, $N_f= 2$, and $\Lambda = 4.8$. The value of $\Lambda$ is chosen to set the critical temperature to be around $S_0 \sim 12$. 

The dyon partition function and interactions are identical to that of the ensemble in the pure $SU(3)$ theory \cite{DeMartini:2021dfi} with two additional terms: a potential arising from the perturbative interactions of the quarks with the holonomy, and quark-induced interactions between $L$ and $\bar{L}$ dyons. Here we provide a brief description of the partition function, focusing on the new terms. 

The partition function of the dyon ensemble is separated into two parts as $Z= Z_0 Z_{int}$, where $Z_0$ is the non-interacting terms and $Z_{int}$ contains all of the dyon interactions. The partition function is
\begin{equation}
\begin{aligned}
	Z_0= &e^{\tilde{V}_3 (V_{GPY} + V_{quark})}\sum_{N_{M1},N_L,N_{M2}} \left(\frac{1}{N_{M1}!} (\tilde{V}_3 d_{\nu})^{N_{M1}} \right)^2 \\
	& \times \left(\frac{1}{N_L!} (\tilde{V}_3 d_{1-2\nu})^{N_L} \right)^2 \left(\frac{1}{N_{M2}!} (\tilde{V}_3 d_{\nu})^{N_{M2}} \right)^2 ,
\end{aligned}
\end{equation}
 where $d_{\nu}$ is the weight of an individual dyon with holonomy $\nu$ \cite{Diakonov:2004jn}
 \begin{equation}
 d_{\nu} = \frac{\Lambda}{4\pi} S_0^2 e^{-S_0 \nu} \nu^{\frac{8 \nu}{3} - 1}
 \label{weight}.
 \end{equation}

The potentials in the partition function are the perturbative potentials of the quarks and gluons with the holonomy. The gluons experience the well-known Gross-Pisarski-Yaffe potential $V_{GPY}$ \cite{Gross:1980br}, which in $SU(3)$ is 
   \begin{equation}
   \frac{V_{GPY}}{\tilde{V}_3 T} = \frac{4 \pi ^2}{3}(2(\nu(1-\nu))^2 + (2\nu(1-2\nu))^2).
   \end{equation}
   
\begin{figure}[h]
      	\includegraphics[width=0.85\linewidth]{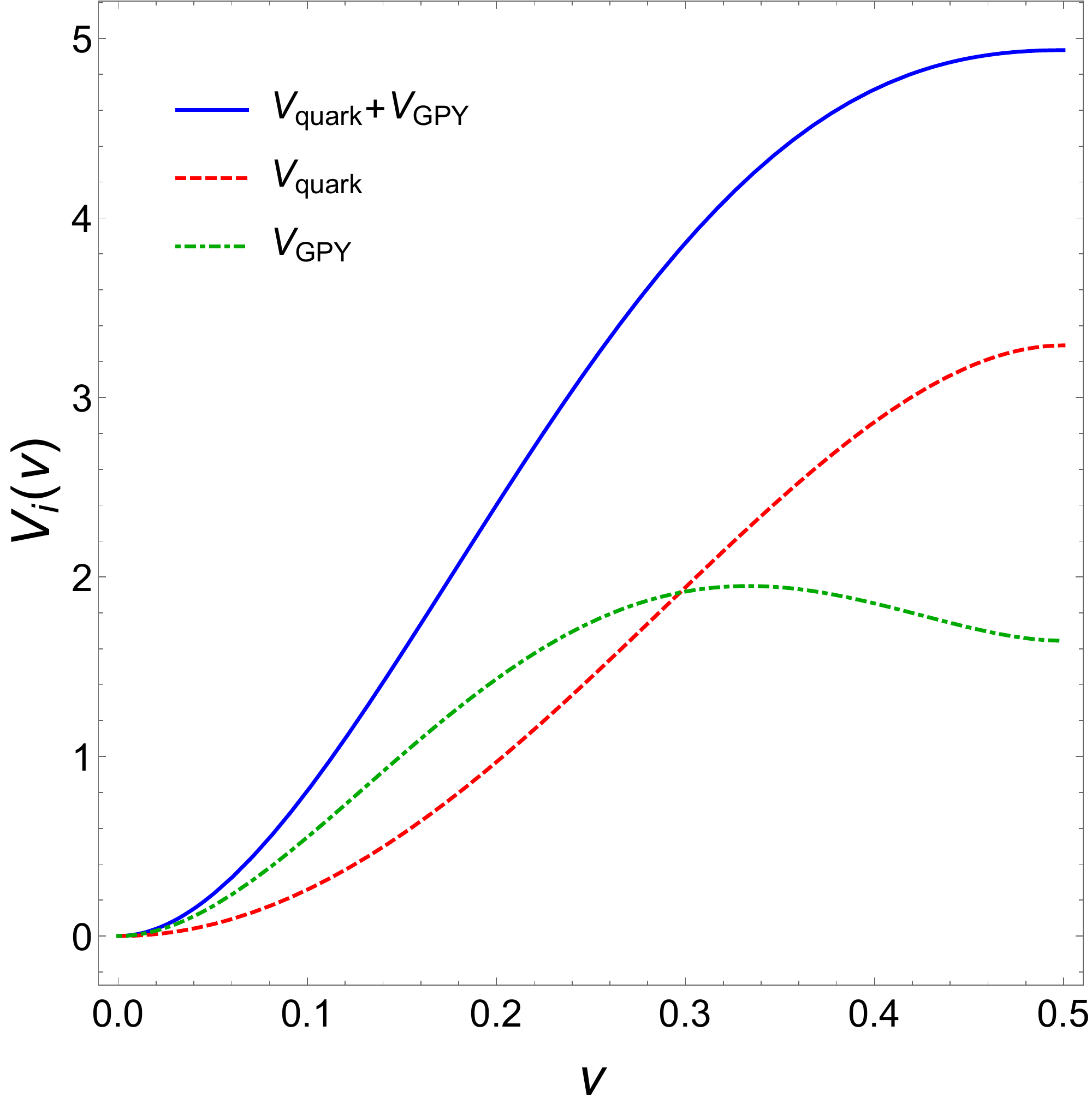}
      	\caption{(Color online) Holonomy dependence of the perturbative $SU(3)$ quark and gluon potentials and their sum for $N_f=2$ and $\tilde{V}_3 T =1$. }
      	\label{fig_pot}
      \end{figure}

The first term new to this work is the second potential $V_{quark}$, which is the analogous interaction of $N_f$ flavors of massless quarks with the holonomy \cite{Weiss:1981ev,Fukushima:2017csk}. For $SU(3)$, it has the form

\begin{equation}
\frac{V_{quark}}{\tilde{V}_3 T} = -N_f \frac{4 \pi^2}{3}(2 \nu^4 - \nu^2).
\label{eq_pot}
\end{equation}
Both of these potentials favor the deconfining holonomy $\nu = 0$. Fig. \ref{fig_pot} shows the contributions of both terms to the free energy density. 

\subsection{Dyon interactions}

Within the temperature range of interest near the critical temperature $T_c$, the dyons are strongly interacting. These interactions contribute significantly to the partition function and their computation is nontrivial, requiring Monte-Carlo integration over the $3N_D$-dimensional space of the dyons' collective coordinates. The interaction terms of the partition function can be written in the form 

   \begin{equation}
   \begin{aligned}
   Z_{int} = &\frac{1}{\tilde{V}_3^{(4N_{M}+2N_L)}} \int Dx \det{(G)} \det{(\bar{G})}  e^{-\Delta S_{cl}} \\
   & \times \big(\prod \lambda_i \big)^{N_f}
   \end{aligned}
   \end{equation}
Here the interactions are separated into three parts: the classical binary interactions of the dyons $\Delta S_{cl}$, the one-loop fluctuation determinants $\det{(G)}$ and $\det{(\bar{G})}$, and the eigenvalues $\lambda_i$ of the Dirac operator due to the inclusion of $N_f$ flavors of dynamical massless quarks. 

The classical and one-loop interactions included are the same as those in the pure $SU(3)$ theory. At large distances, the dyon interactions are Coulomb-like. We use the parameterized form
\begin{equation}
   \Delta S_{cl}^{d \bar{d}}= -\frac{S_0 C_{ij}}{2\pi}(\frac{1}{rT}-2.75 \pi \sqrt{\nu_i \nu_j} e^{-1.408 \pi \sqrt{\nu_i \nu_j} r T}),
\end{equation}
between dyons $i$ and $j$. The coefficient $C_{ij}$ is $-2$ for dyon-antidyon pairs of the same type, $1$ for dyon-antidyon pairs of different types, and $0$ for dyon-dyon or antidyon-antidyon pairs. At distances shorter than the core size $x_0 = 2 \pi \nu_i r_0 T$, dyons of the same type, regardless of duality, experience a repulsive core of the form 
\begin{equation}
   \Delta S_{cl}^{core} = \frac{\nu V_0}{1+e^{2 \pi \nu  T (r - r_0)}}.
\end{equation}

The volume metrics ($G$ for the dyons and $\bar{G}$ for the antidyons) are the Diakonov determinants \cite{Diakonov:2009jq}, each with the same form for the elements between the $i$-th dyon of type $m$ and the $j$-th dyon of type $n$
   \begin{equation}
   \begin{aligned}
   G_{im,jn} ={} & \delta_{ij} \delta_{mn} (4\pi \nu _m - \sum_{k\ne i} \frac{2}{T|r_{i,m} - r_{k,m}|}\\
   & + \sum_{k} \frac{1}{T|r_{i,m}-r_{k,p \ne m}|}) \\
   &+ \frac{2\delta _{mn}}{T|r_{i,m}-r_{j,n}|}-\frac{1- \delta_{mn} }{T|r_{i,m}-r_{j,n}|}.
   \end{aligned}
   \end{equation}  
   
The other  term in the interactions is fermionic determinant, not included in our previous work. It can be written as the product of the eigenvalues of the Dirac operator $ \det (\slashed{D}) = \prod \lambda_i $. Its form is discussed in detail in the next section.

Additionally all Coulomb-like terms in the classical and one-loop interactions are regulated by an electric Debye screening term $e^{M_D r T}$. There are three phenomenological parameters of the model which at present are not known from first principles, $V_0$, $x_0$, and $M_D$. We use the values $V_0 = 40$, $x_0 = 1.8$, and $M_D = 1.5$ in this work.   

The contribution of the interaction potential to the free energy density is computed via the standard integration over a dummy parameter $\lambda$,
\begin{equation}
\Delta f = \frac{1}{\tilde{V}_3} \int_0^1 d\lambda \langle \Delta S \rangle.
\end{equation}    
In total, the free energy density for an ensemble with specified input parameters is 
\begin{equation}
\begin{aligned}
f&(T,\nu,n_M,n_L) = - N_f \frac{4 \pi^2}{3}(2\nu^4-\nu^2) \\
&+ \frac{4 \pi ^2}{3}(2(\nu(1-\nu))^2 + (2\nu(1-2\nu))^2) \\
& -4n_M \ln \left[\frac{d_{\nu}e}{n_M}\right]- 2n_L \ln \left[\frac{d_{1-2\nu}e}{n_L}\right] \\ 
& + \frac{\ln(8\pi^3 N_M^2 N_L)}{\tilde{V}_3} +\Delta f, \\
\end{aligned}
\label{eqfree}
\end{equation} 
where $Z_0$ has been expanded with Stirling's approximation to three terms.
The central goal of this work is then to compute $f$ for a range of input parameters and determine the location of the minimum for each value of $T$, thereby determining the physical properties of the ensemble as functions solely of the temperature. 

\section{Quark-Induced Interactions} \label{sec_quark}

\subsection{Fermionic determinant}

The main result of incorporating dynamical quarks into the gauge theory is the inclusion of the fermionic determinant in the partition function. In the context of the instanton-dyon ensemble, the fermionic determinant is approximated by considering only the subspace of quark states spanned by the zero modes. This approximation results in the 'hopping matrix'
\begin{equation}
(\det (\slashed{D} + m_a))^{N_f} \approx (\det(\hat{T}))^{N_f},
\end{equation}
assuming equal masses for all quark flavors $a$. Here we have removed a factor of $i$ from the l.h.s. of the equation in order to write $\hat{T}$ as a purely real matrix. 

For physical (antiperiodic) quarks, the right-handed zero modes are localized on the $L$ dyons and the left-handed zero modes are localized on the $\bar{L}$ dyons. Individual elements of the hopping matrix can be interpreted as the 'hopping amplitude' for a quark going between $L$ dyon $i$ and $\bar{L}$ dyon $j$. These elements are then given by
\begin{equation}
T_{ij} = \langle i| \slashed{D} + m |j \rangle = \int d^4x \psi_i ^{\dagger} (x-x_i) (\slashed{D}+m) \psi_j(x-x_j).
\end{equation}

If one approximates the total gauge field as the sum of the fields of two dyons, then the covariant derivative can be reduced to an ordinary derivative by the zero mode equations of motion. This hopping amplitude only has a nonzero contribution from $\slashed{D}$ when the zero modes have opposite chirality (i.e. hopping within $L\bar{L}$ pairs, but not $LL$ or $\bar{L}\bar{L}$) and because the zero modes are orthogonal, the mass term only contributes along the diagonal. Thus, the hopping matrix takes on the simple form
\begin{equation}
\hat{T} = \begin{pmatrix}
m\delta _{ij} & T_{ij} \\
-T_{ji} & m\delta _{ij} \\
\end{pmatrix}.
\label{eq_mat}
\end{equation}

\begin{figure}[h]
	\includegraphics[width=0.85\linewidth]{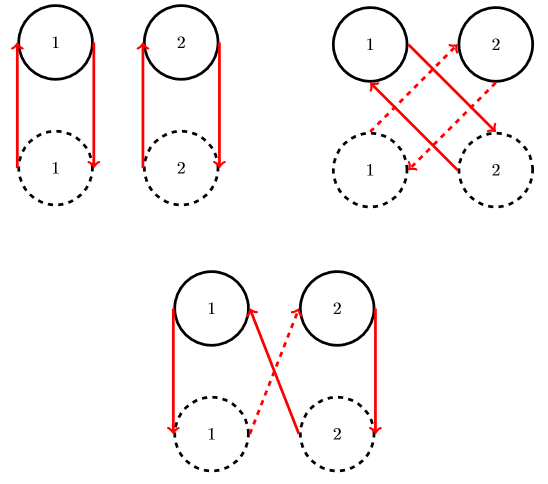}
	\caption{(Color online) The set of diagrams corresponding to $\det (\hat{T})$ for the system of two $L$ dyons (solid circles) and two $\bar{L}$ antidyons (dashed circles). Red arrows represent quark hoppings between dyons. Each diagram corresponds to a term in Eq. (\ref{eq_det}).}
	\label{fig_loop}
\end{figure}

The hopping matrix $\hat{T}$ is a $2N_L \times 2N_L$ antisymmetric matrix in the case of massless quarks which we consider in most of this work. The fermionic determinant may be interpreted as a sum of all closed loops of the quarks hopping between dyon-antidyon pairs. As an explicit example, let us consider the case of two $L$ dyons $1$ and $2$ and two $\bar{L}$ dyons $\bar{1}$ and $\bar{2}$. In this case the determinant of the hopping matrix is 
\begin{equation}
	\det(\hat{T}) = T_{11}^2 T_{22}^2 + T_{12}^2 T_{21}^2 - 2 T_{11}T_{21}T_{22}T_{12}.
	\label{eq_det}
\end{equation}
The first two terms correspond to the two possible two-loop diagrams in which each $L$ dyon forms a closed loop with a single $\bar{L}$ dyon each (the upper diagrams in Fig. \ref{fig_loop}). The last term represents the one-loop diagram in which the quark hops between all four dyons (the bottom diagram in Fig. \ref{fig_loop}). For arbitrary number of $L$-dyons $N_L$, there are $N_L!$ terms in the determinant of the hopping matrix. (Of course, numerically we evaluate the determinant
by a more efficient algorithm.)  

Whether the determinant is dominated by paths involving single pairs ('instanton molecules') or paths of many dyons ('instanton polymers') determines the state of the chiral symmetry. Consider the case in which the dyons are arranged into well-separated $L\bar{L}$ pairs. In this case, the determinant is dominated by $T_{ii}$ and $T_{ij} \rightarrow 0$ for $i \ne j$. In this case the eigenvalues of the matrix are $\lambda \approx \pm T_{ii}$. In this configuration, the eigenvalues are then all large for the nearby dyon-antidyon pairs, suppressing the density of near-zero eigenvalues, leading to the disappearance of the quark condensate. It is in the collectivized 'polymer' regime, where there are many $T_{ij}$'s of comparable magnitude which nearly cancel each other in the eigenvalues, which possesses a non-vanishing density of eigenvalues near zero. 

\subsection{Parameterization of the hopping matrix elements}
 
 The general form of the quark zero modes on the $SU(N)$ dyon gauge fields was first given in Ref. \cite{GarciaPerez:1999ux}. An explicit form for the zero mode density, in terms of all dyon coordinates is far too complicated to be of a practical use. (See Appendix A for a discussion of the general solution.) Instead, we start here with the form of the zero mode for a lone $L$ dyon\footnote{In the equations in this section we explicitly restore factors of $2\pi$ and $T$ which are suppressed in other works.}:
 \begin{equation}
 \psi_a^A = \sqrt{2 \pi} \frac{\bar{\nu} \tanh (\pi \bar{\nu} r T)}{\sqrt{r T \sinh (2 \pi \bar{\nu} r T)}} e^{i \pi \tau T} \epsilon_a^A,
 \label{eq_zero}
 \end{equation}  
 where the $\epsilon$ symbol contains the color and spin structure. The left-handed zero mode on the $\bar{L}$ dyon is found by changing the spin of the quark in the $\epsilon$ symbol. The spatial structure of the wavefunction is the same for all $SU(N)$. In this limit, the zero mode density has no Euclidean time dependence. The effect of nearby $M_i$ dyons is to \textit{destructively} interfere with the gauge field of the $L$ dyon localizing the zero mode in both space and time. For some observables, such as hadronic correlation functions \cite{Larsen:2017sqm}, this inon-binary forces seem to be necessary to achieve reasonable results.
 
 In some preliminary computations for this work, an \textit{ad-hoc} approach to including the effects of interference was considered by appending a term estimating the localization effects to the wavefunction in Eq. (\ref{eq_zero}). 
 It was found that these terms led to only a modest modification of the quark-induced interactions. Thus we do not include such effects in out parameterization of the hopping matrix. 
 
 Computing $T_{ij}$ requires numerical integration, and instead a parameterization of it must be used. For a detailed derivation of $T_{ij}$ see Ref. \cite{LopezRuiz:2019hvh}. We use the same parameterization as in the $SU(2)$ work \cite{Larsen:2015tso}, known there as 'Parametrization A,' 
 \begin{equation}
 T_{ij} = \bar{\nu} c' \exp(-\sqrt{11.2+(\pi \bar{\nu} r T)^2}).
 \end{equation} 
 The magnetically-charged dyons have Dirac strings which are, in principle, pure gauge artifacts. However, the use of the sum ansatz in combining the two gauge fields introduces gauge-dependent factors in the zero modes. This leaves ambiguity in the overall normalization, handled here by $c'$ which we treat as a tunable parameter of the model and have chosen to use $\ln(c') = 4.45$. 
 
 The fermionic determinant adds an effective potential to the ensemble of the form 
 \begin{equation}
 \Delta S_{quarks} = -N_f \ln (\det (\hat{T})).
 \end{equation}
This induces complicated many-body interactions between all $L$ and $\bar{L}$ dyons, but at the simplest level is an attractive force within $L\bar{L}$ pairs. 

\section{ Simulations and Analysis} \label{sec_tech}

The simulation setup and analysis follows much of what was done in the previous work (see Section III of Ref. \cite{DeMartini:2021dfi}). The free energy density of the dyon ensembles are computed via Monte-Carlo integration using the standard Metropolis algorithm. For each set of input parameters the simulation is run with 10 values of the dummy parameter $\lambda = 0.1,...,1$. Every dyon position is updated five times between samplings and 2000 configurations are sampled for each value of $\lambda$. 

Each simulation is run at fixed dyon number $N_D =120$ (60 dyons and 60 antidyons). The densities of the dyons are controlled by varying the length $L$ of the sides of the simulation box and the relative number of each type of dyon. Periodic (spatial) boundary conditions are imposed by a set of 26 image boxes placed around the main simulation box. Because of the large cost of computing the determinants of the matrices, only dyon interactions within a local box of length $L$ centered on the dyon whose position is being updated are computed at each update step. Compared to the pure $SU(3)$ case, the only addition to the Metropolis update step is computing the fermionic determinant. Because it only involves the positions of the $L$ and $\bar{L}$ dyons, it is much smaller than the Diakonov determinant and does not increase the computational cost significantly. This is in contrast to lattice simulations, where the fermionic determinant is typically the most computationally-expensive task. 

For any finite number of images there are dyons near the faces of the total simulation box which feel unphysical effects of the boundaries. In the pure $SU(3)$ case, where all interactions were exponentially suppressed by the Debye mass, these finite-volume effects were only a percent-level correction to the free energy. The quark-induced interactions are linear at long distances and must be treated more carefully. When computing the potential from these interactions, for each pair of dyons $i$ and $j$ one should consider only the image of $j$ which minimizes the distance between the pair. This ensures that the short-distance (large-eigenvalue) interactions of dyons near the faces of the boxes aren't excluded while long-distance (small-eigenvalue) of dyons on the opposite sides of the boxes aren't over-counted. This introduces an effective cutoff distance $L/2$ considered in the hoping matrix which is remedied when taking the infinite-volume limit. Without using this technique, the eigenvalue distribution has a large number of unphysical, arbitrarily-small eigenvalues.

\begin{table}[h]
	\caption{Ranges of the input parameters used for the main simulation runs.}
	\begin{tabular}{|c|c|c|c|}
		\hline
		& min. & max. & step size \\
		\hline
		$S_0$	& 8 & 14.5 & 0.5 \\
		\hline
		$\nu$ & 0.1933 & 0.3533 & 0.01 \\
		\hline
		$n_M$	& 0.12 & 0.45 & 0.015 \\
		\hline
		$N_M$ & 19 & 28 & 1 \\
		\hline	
	\end{tabular}
	\label{tabparam}
\end{table}

The physical parameters of the dyon ensemble are determined by a fit near the free energy minimum in the space of the input parameters for each value of $S_0$. Once the minima are found, additional simulations are run with the physical parameters generating 60,000 configurations to compute the eigenvalue distributions with better statistics. In order to more accurately represent both fitted densities simultaneously, the dyon number of these runs is allowed to vary slightly with $120 \le N_D \le 128$. Additionally, in order to study finite-volume effects, 30,000 configurations are generated with $2N_D$ dyons and 20,000 configurations are generated with $3N_D$ dyons at the physical parameters for each temperature (thus the number of eigenvalues computed is the same for all three ensemble sizes). 
 
\section{The Polyakov Loop and Deconfinement} \label{sec_poly}

\begin{figure}[h]
	\includegraphics[width=0.85\linewidth]{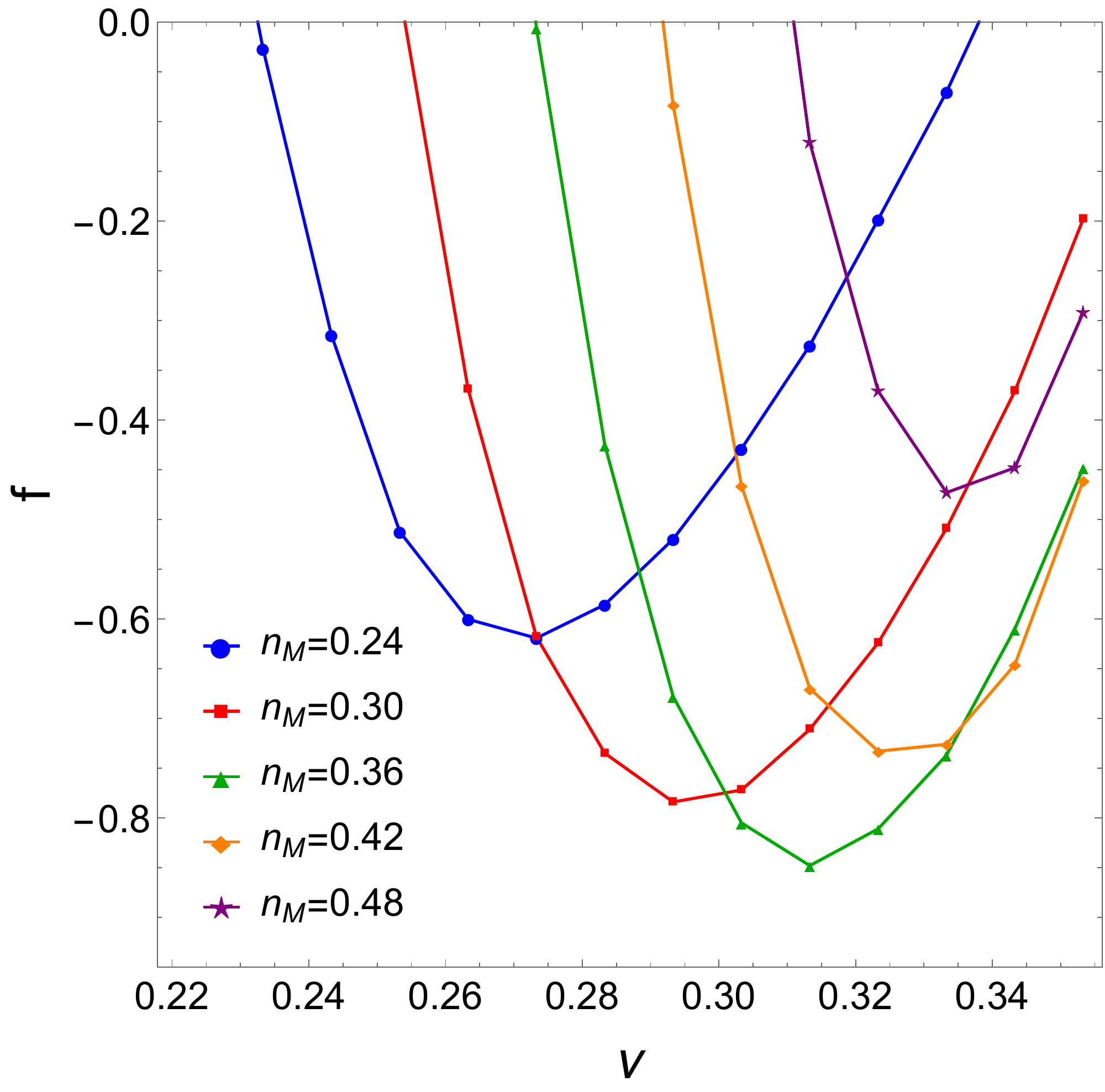}
	\caption{(Color online) Holonomy dependence of the free energy density for different values of the $M$-dyon density $n_M$ with $S_0=9.0$ and $n_L = n_M/1.643$. This ratio of densities is the closest to the fitted value that can be achieved with $N_D=120$ dyons. Error bars not shown for readability.}
	\label{fig_struct}
\end{figure}

The physical properties of the dyon ensemble are determined by the location of the free energy minimum in the space of input parameters. The holonomy- and density-dependence of the potential determines the Polyakov loop and the nature of the deconfinement phase transition. Fig. \ref{fig_struct} gives an example of the holonomy potential. The same general features are seen here as in the pure $SU(3)$ case, namely that at high densities the minimum is located in the confining phase ($\nu = 1/3$) as the dyon interactions dominate, while as the densities are reduced, the minima are pushed to lower values of $\nu$, driven by the perturbative potentials. It is the location of the global minimum, at some intermediate densities, that represents the physical value of the holonomy in the thermodynamic limit. Crucially, however is the fact that the holonomy now varies smoothly as a function of temperature. Unlike the pure $SU(3)$ case, there are not two nearly-degenerate minima near $T_c$ (see Fig. 4 of Ref. \cite{DeMartini:2021dfi}).  

In QCD the deconfinement transition is a smooth crossover occurring almost simultaneously with the chiral transition, $T_{deconf} \sim T_c$ \cite{Bazavov:2009zn,Borsanyi:2012rr}. In the $N_f=2$ massless case we consider, evidence is less conclusive. Recently much progress has been made, in particular by the Bielefeld group \cite{Ding:2018auz,HotQCD:2019xnw,Clarke:2019tzf,Ding:2020eql,Kaczmarek:2020sif,Clarke:2020htu}, on studying the phase transition in $(2+1)-$QCD on the lattice in the chiral limit -- QCD with massless up and down quarks and a physical strange quark $m_s \approx 95$ MeV. Of note is the fact that the location and form of the deconfinement transition is very sensitive to the light quark masses. 

The order of the phase transition in the chiral limit is dependent on the state of the $U(1)_A$ symmetry breaking near the chiral restoration temperature. If the $U(1)_A$ breaking remains significant at these temperatures, as is expected, the phase transitions are expected to be second order belonging to the $O(4)$ universality class, for large enough values of the strange quark mass $m_s$ \cite{Pisarski:1983ms,Butti:2003nu}. If the $U(1)_A$ breaking is small near $T_c$, the transition may be first order, although the recent lattice results disfavor that scenario. The phase diagram of the $O(4)$ class is characterized by the temperature $T$ and external field $H$. In the case of the gauge theory, the role of the external field is played by the light quark masses, meaning we look at the $H=0$ line with the massless quarks. 

\begin{figure}[h]
	\includegraphics[width=0.85\linewidth]{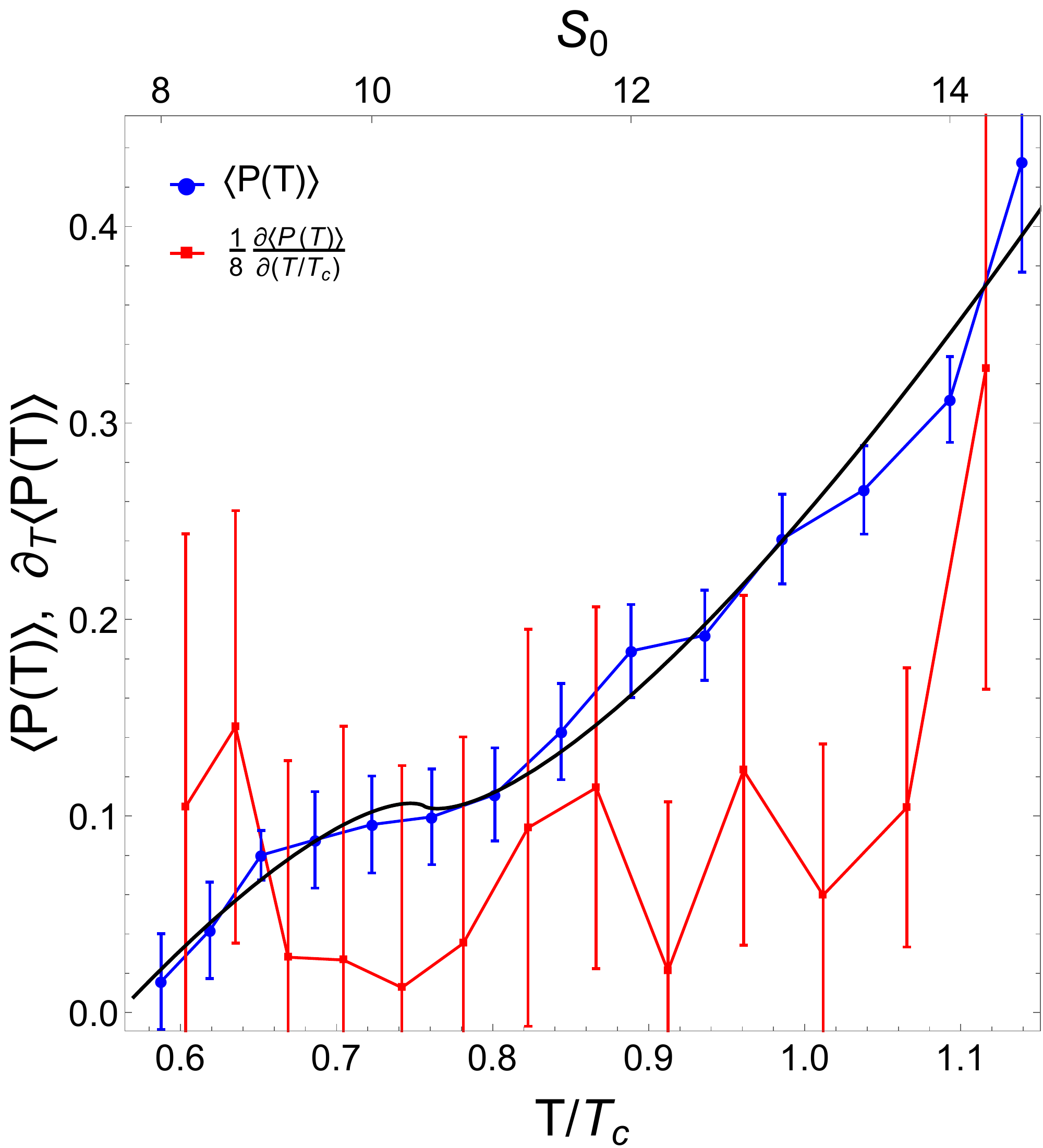}
	\caption{ (Color online) Temperature dependence of the average Polyakov loop and its (scaled) temperature derivative in the dyon ensemble. Derivative is computed from the differences in consecutive data points. Solid curve shows the fit to the form of the second-order transition in the $O(4)$ universality class (\ref{eq_pfit}).}
	\label{fig_poly}
\end{figure}

We then fit the data for $\langle P(T) \rangle$ a the form inspired by Ref. \cite{Clarke:2020htu}.
\begin{equation}
	\langle P(T) \rangle = \exp \big( -a_0 - t(a_1 + A |t|^{-\alpha}) \big),
	\label{eq_pfit}
\end{equation}

where $t= (T-T_{deconf})/T_{deconf}$ and $\alpha=2-\beta (1+\delta)$ is the hyperscaling variable with the values of the $O(4)$ class $\beta=0.380$, $\delta=4.824$, and $\alpha= -0.2131$ \cite{Engels:2003nq,Engels:2011km}. From this fit ($\chi^2 = 0.0257$) we find the critical temperature of the deconfinement transition $S_0(T_{deconf})=10.44 \pm 0.29$. We also constrain the signs of the fit parameters to match those determined from the lattice data $a_0, \, A>0$, $a_1<0$. Removing these constraints or treating $\alpha$ as a fit parameter results in comparably-good fits with very different $T_{deconf}$ values. For example, the $O(2)$ universality class is qualitatively similar but $\alpha=-0.0172$ \cite{Hasenbusch:1999cc,Engels:2000xw} is an order of magnitude different from the $O(4)$ value. The dyon data shows slightly better agreement to that fit ($\chi^2= 0.0178$), so we do not claim that our data supports $O(4)$ over other $O(N)$ universality classes, but merely that it is \textit{compatible with} the expected $O(4)$ behavior. Because of the multiple potential fits, one should consider the determination of $T_{deconf}$ to have larger uncertainties than those given by any individual fit. 

In QCD, with no exact chiral  or $\mathbb{Z}_3$ symmetries,
 the transitions are an analytic crossover. 
 The psuedocritical temperature is defined by the inflection point of the curve, where the derivative with respect to the temperature has a maximum, since there is no universal scaling expected near the transition from which a proper critical temperature could be determined via a fit. Lattice QCD studies find that, for the Polyakov loop, the peak is broad and hard to accurately define. We determine the derivative of $\langle P(T) \rangle$ by simply computing the slopes between points. We see in Fig. \ref{fig_poly} that no distinct peak can be effectively determined due to the large uncertainties.    

\begin{figure}[h]
	\includegraphics[width=0.85\linewidth]{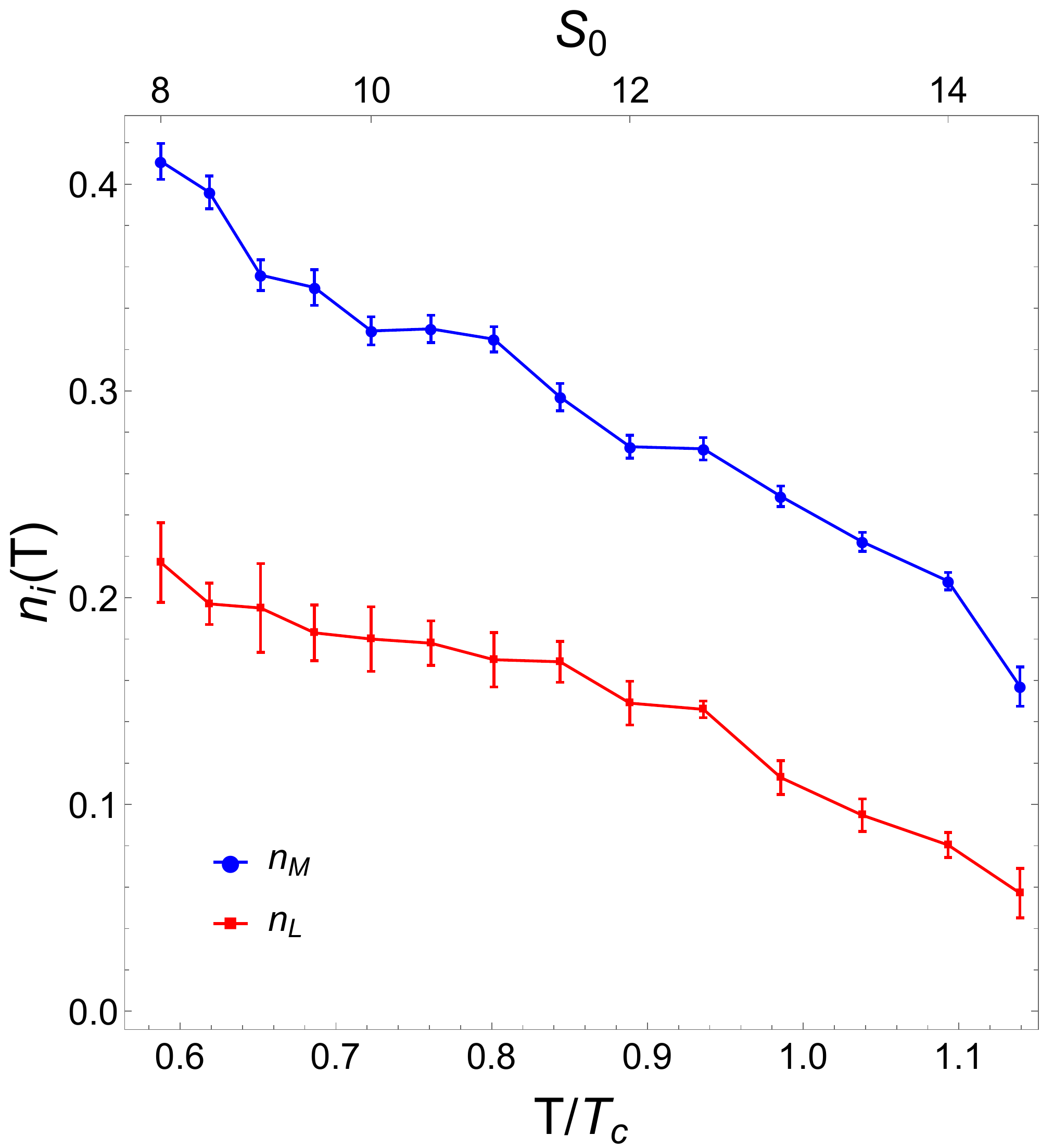}
	\caption{(Color online) Temperature dependence of the (dimensionless) densities for $L$- and $M_i$-type dyons. }
	\label{fig_dens}
\end{figure}

The dyon densities, both shown in Fig. \ref{fig_dens}, decrease as the temperature rises. Unlike the pure $SU(3)$ case, the $L$-dyon density remains smooth around $T_c$. Even in the confined phase, when the dyon actions and sizes are equal, the densities are not due to the quark-induced interactions breaking the symmetry between the dyon types (or more generally, breaking $\mathbb{Z}_3$ symmetry). The ratio of the densities at low $T$ is directly sensitive to our choice of $c'$.

\section{ Chiral Symmetry Breaking} \label{sec_chiral}

\subsection{Eigenvalue distribution of the Dirac operator}

\begin{figure*}
	\includegraphics[width=0.95\linewidth]{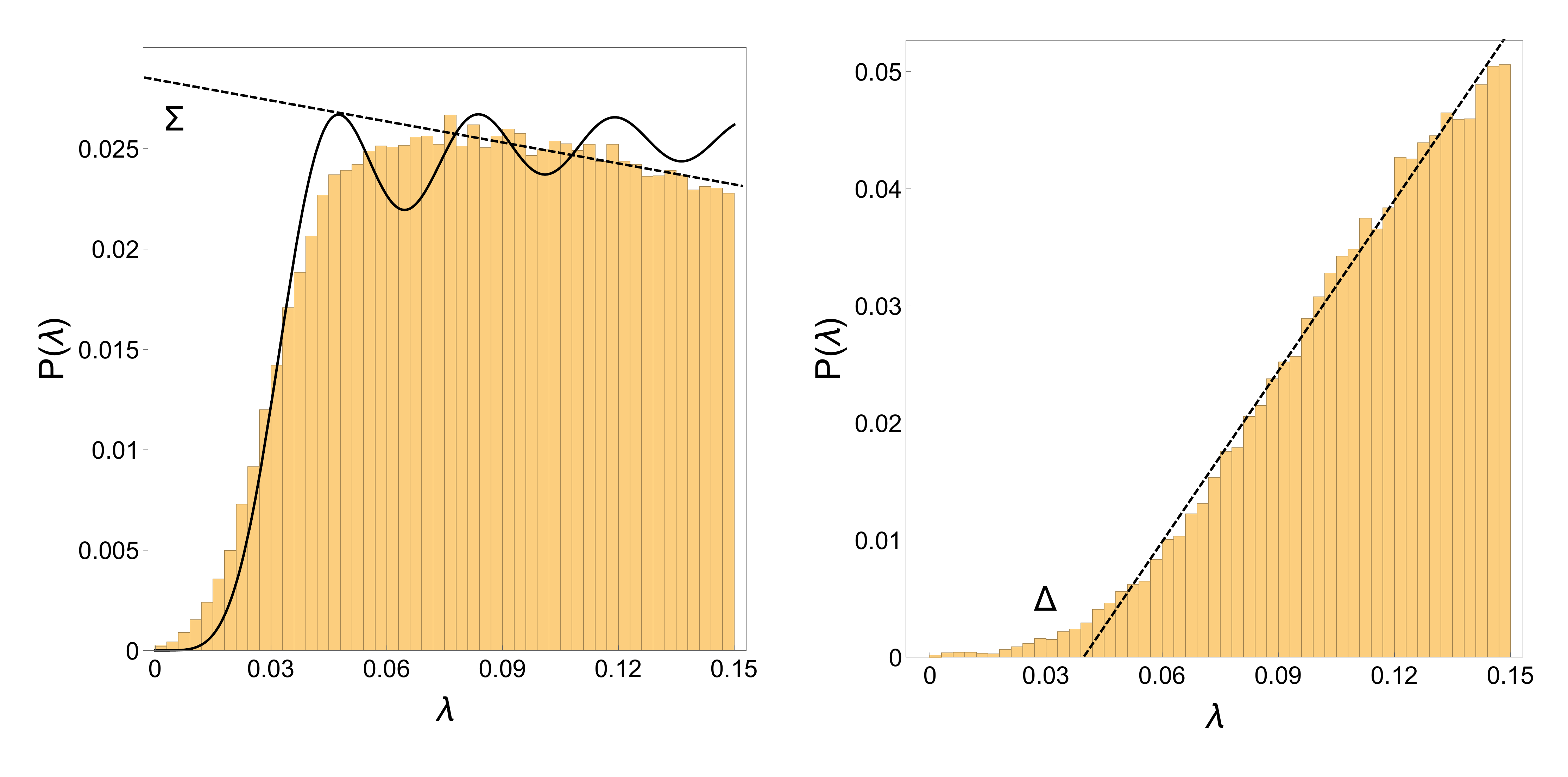}
	\caption{Normalized probability distribution of eigenvalues in the near-zero-mode zone for both the broken and restored phases. Dashed lines show fits to the linear regions of the distributions. Left: $S_0 = 8$, the y-intercept of the fit is proportional to the chiral condensate $\Sigma$, the solid curve is the fit to the random-matrix theory results (\ref{eq_rmt}), Right: $S_0 =14$, the x-intercept of the fit is the eigenvalue gap $\Delta$. Note that the linear fit to determine the y-intercept on the left plot is not used in the analysis and is merely illustrative.}
	\label{fig_hist} 
\end{figure*}

The breaking of chiral symmetry is associated with the existence of a nonzero quark condensate $\langle \bar{q}q \rangle$ generated by nonperturbative effects at low temperatures. The quark condensate is related to the zero eigenvalues of the Dirac operator by the Banks-Casher relation \cite{Banks:1979yr}

\begin{equation}
\Sigma = |\langle \bar{q} q \rangle| = \lim\limits_{\lambda \rightarrow 0}\lim\limits_{V \rightarrow \infty} \pi \rho(\lambda),
\label{eq_bcr}
\end{equation}
where $\rho(\lambda)$ is the spectral density of the Dirac operator. 

With the hopping matrix being antisymmetric, its spectrum is symmetric in the sign of the eigenvalue, $\rho(\lambda) = \rho(-\lambda)$. For simplicity, we only show the positive spectrum in the plots. Fig. \ref{fig_hist} shows examples of the spectrum in both phases. 

The smallest eigenvalues correspond to collectivized modes with the zero modes of many dyons overlapping. For a system of finite size, the eigenvalues below some value $\lambda_{min}$ are suppressed with $\lambda_{min} \propto 1/V$, meaning that regardless of which phase the system is in, a finite system will always have zero eigenvalue density near zero. 

In Fig. \ref{fig_hist} (left), the steep decrease in eigenvalues below $\lambda \sim 0.06$ is a result of the finite size of the system. The dashed line extrapolating to $\lambda=0$ gives an estimate of the spectral density in the absence of such effects. In the restored phase (Fig. \ref{fig_hist} (right)), the there is a larger range of decreasing eigenvalues. The decrease in density near zero reflects a real absence of collectivization rather than finite volume effects. This can't be seen from a single eigenvalue distribution alone, and distinguishing the two phases requires analyzing the distributions for multiple volumes. 

\subsection{Infinite-volume extrapolation and results}

We locate the chiral phase transition by two different methods. The first is by extracting the chiral condensate $\Sigma(T)$ by extrapolating the small eigenvalue distribution to infinite volume using results from random-matrix theory. The vanishing of the condensate is related to the psuedocritical temperature $T_c$. The other way is to use the eigenvalue gap $\Delta(T)$ by fitting the smallest eigenvalues to a linear function (similar to Fig. \ref{fig_hist} (right)). Above a temperature $T_{gap}$, the restoration of chiral symmetry leads to a finite size of the smallest eigenvalues, meaning the lowest excitations have finite mass.  

The mesoscopic volume scaling of the near-zero eigenvalues can be understood from random-matrix theory, which for $N_c=3$, $N_f=2$, gives a Dirac eigenvalue spectrum of the form \cite{Verbaarschot:1993pm}

\begin{equation}
	\rho(z)= V \Sigma_2 \left( \frac{z}{2} (J_2^2(z)-J_1(z)J_3(z)) \right),
	\label{eq_rmt}
\end{equation}
where $z= \lambda V \Sigma_1$ and $J_n$ are the Bessel functions. Here $\Sigma_1$ is the scaling factor and $\Sigma_2$ is the overall normalization factor. In the infinite volume limit $\rho(0) \rightarrow V \Sigma_2$. 

Both of the factors $\Sigma_1$ and $\Sigma_2$ are related to different physics with different volume-dependence. In the case of the dense, low-temperate ensemble the eigenvalue distribution should be dominated by the collectivized modes. In this case, increasing the volume should reduce the region of suppressed eigenvalues by the same factor and $\Sigma_1 \propto V \propto N_D$. On the opposite end, when the ensemble is dilute and comprised of dyon-antidyon pairs, $\Sigma_1$ is independent of the system size. Of course, in the region near $T_c$, the ensemble is a mixture of both components and we must interpolate between the two.

A fit to the distribution gives two parameters per ensemble size. To extract the infinite-volume value of the condensate we use an interpolating function to determine how much of the region of smallest eigenvalue is real or a finite volume effect. We use the function

\begin{equation}
	\Sigma = \Sigma_2 \left( \frac{2\Sigma_1^{3V}}{\Sigma_1^{2V}} -2 \right) \left( \frac{\Sigma_1^{2V}}{\Sigma_1^V} - 1  \right),
	\label{eq_int}
\end{equation}
where $\Sigma_2=(\Sigma_2^{V} + \Sigma_2^{2V} + \Sigma_2^{3V})/3$ is the overall scale and each term in parenthesis we call a scaling factor. 

The philosophy of these scaling functions is as follows: each scaling factor is linear in the ratios of $\Sigma_1^i$ and chosen such that it gives 0 or 1 in the opposite cases described above. In the case where the lowest-eigenvalue portion of the spectrum is real and doesn't change with the volume, the scaling factors are 0, and thus the condensate is 0 as there is a finite eigenvalue gap. In the case where the suppressed region is entirely due to finite volume effects, the factors should scale with $V$ and $\Sigma_1^{3V} = 3/2 \Sigma_2^{2V}$, $\Sigma_1^{2V} = 2 \Sigma_2^{V}$ the scale factors are 1 meaning the overall value of the condensate remains in the infinite-volume limit $\Sigma= \Sigma_2$. Additionally we enforce an upper bound of 1 on each scaling factor in cases where the values scale faster than $V$. Two scale factors are used to scale between the three volumes. Alternative functions using just two of the volumes are discussed in Appendix \ref{app_extra}. 

Along the $H=0$ line, the chiral condensate (analogous to the magnetization $M$ in the $O(4)$ spin model) takes the form
\begin{equation}
	\Sigma(T)= \left\{
	\begin{array}{ll}
		C(T_c-T)^{\beta} & \text{if} \,\,\, T<T_c \\
		0 & \text{if} \,\,\, T \geq T_c \\
	\end{array}
	\right.
	\label{eq_cond}
\end{equation}
where C is some (non-universal) constant. We qualitatively compare this form to our results in Fig. \ref{fig_cond}. As expected, we see a very rapid drop in the condensate to zero just below $T_c$. More data points just above and below $T_c$ would be needed to get a more accurate fit to the data. Going to the lowest temperatures we do not see the continued increase in $\Sigma$ expected by Eq. (\ref{eq_cond}), however this universal scaling behavior is only applicable in the region near the critical temperature. As with the deconfinement transition, we claim our results \textit{are compatible with} the second-order transition of the $O(4)$ class. Certainly, like the lattice data, our results disfavor the potential first-order transition. 

To determine the eigenvalue gap, a linear fit is performed on the smallest eigenvalues for each of the ensemble sizes. The x-intercept of each fit gives the eigenvalue gap. In order to extract the true value of the gap and remove finite-volume effects, a linear function in $1/V$ is fit to the three gaps for different volumes. The value of the gap at $1/V=0$ gives the true value of the gap $\Delta(T)$ in the absence of the finite-volume suppression of small eigenvalues. The results of this analysis are also seen in Fig. \ref{fig_cond}. 

Below a certain temperature, the gaps are all compatible with zero, indicating a finite density of zero eigenvalues. At $S_0 \sim 13.0-13.5$, a finite gap forms and quickly rises, nearly simultaneously with the steep drop in the value of the condensate. Both of the fitting methods for the condensate $\Sigma(T)$ and gap $\Delta(T)$ give consistent temperatures for the chiral symmetry restoration. From a fit of the condensate to the $O(4)$ form, we get $S_0(T_c) = 13.14 \pm 0.14$, the central value of which sets the relative temperature scale in our plots. With this result, there is still one nonzero data point above $T_c$. The fit shown in the plot provides better agreement with the data points just below $T_c$ and is also more in line with the appearance of a nonzero gap. 

\begin{figure}[h]
	\includegraphics[width=0.85\linewidth]{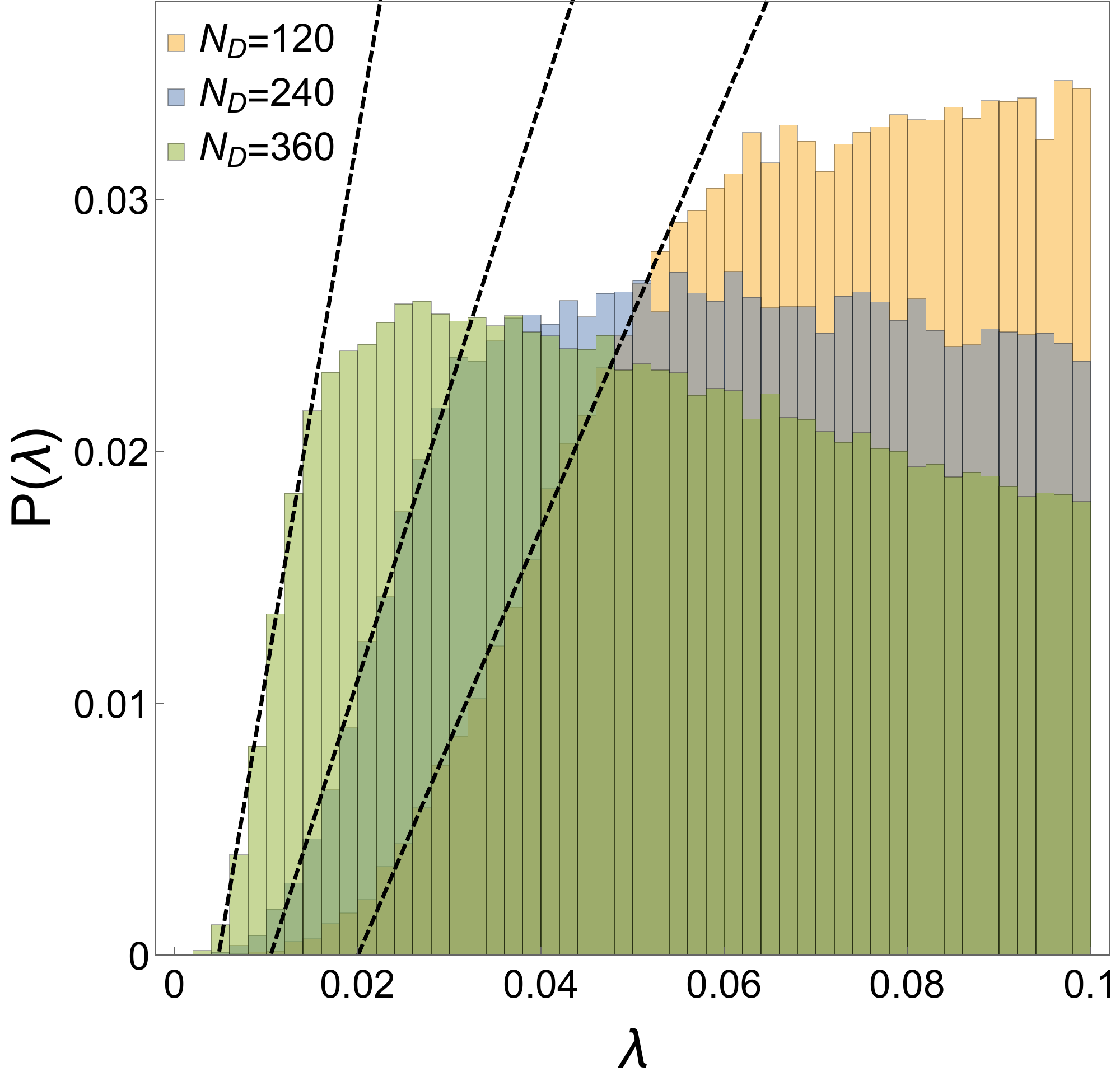}
	\caption{(Color online) Eigenvalue distributions at $S_0=8.5$ for three different ensemble sizes. Dashed lines represent fits to the approximately-linear portion of the distribution near zero. The eigenvalue gaps are given by the x-intercepts of the fits. Note that the relative normalization of the distributions does not affect results.}
	\label{fig_gaps}
\end{figure}

Like deconfinement, in massive QCD the chiral phase transition is an analytic crossover \cite{HotQCD:2018pds}, but is second order in the massless case. 
One of the most interesting results from the $(2+1)-$QCD lattice works \cite{Ding:2018auz,HotQCD:2019xnw,Clarke:2019tzf,Ding:2020eql,Kaczmarek:2020sif,Clarke:2020htu} is that taking the chiral limit and reducing the light quark masses by just a few MeV results in a significantly-reduced transition temperature $T_c$ \cite{Kaczmarek:2020sif,HotQCD:2018pds}.
Critical temperatures for the physical, massive case and the massless limit are
\begin{equation}
	T_c^{QCD}= 156.5\pm1.5 \textrm{  MeV}, \,\,\,\,\, T_c^{2+1} = 132_{-6}^{+3} \textrm{  MeV}.
\end{equation} 

Our results for 
 the $N_f=2$ massless case are shown in Fig. \ref{fig_cond}, in which we show both the quark condensate and the eigenvalue gap values extrapolated to the infinite volume limit.  Both of them indicate
 the same location of the critical temperature, which is finally determined by a fit with
 expected critical exponent (solid curve). Translating our scale to
approximate absolute temperature, we found that our simulations cover the temperature range of about $80$ MeV $< T < 150$ MeV.

\begin{figure}[h]
	\includegraphics[width=0.85\linewidth]{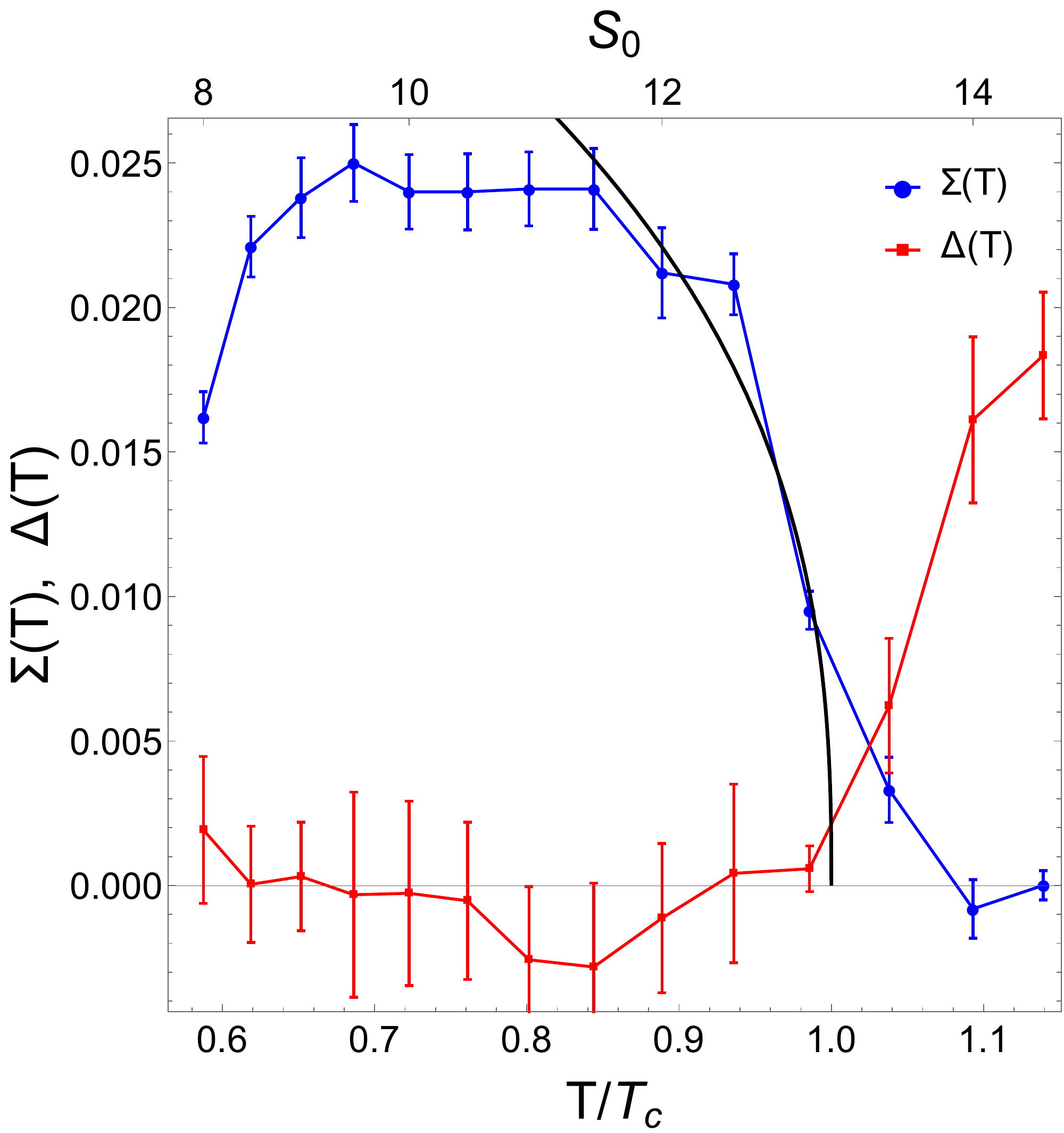}
	\caption{(Color online) The chiral quark condensate $\Sigma(T)$ and the eigenvalue gap $\Delta(T)$ as functions of the temperature. Solid curve is the form of the condensate from the $O(4)$ universality class (\ref{eq_cond}) and is a fit to the four data points just below $T_c$. The overall normalization of $\Sigma(T)$ is arbitrary.} 
	\label{fig_cond}
\end{figure}

\subsection{Effects of nonzero quark mass}

A nonzero quark mass explicitly breaks chiral symmetry in the QCD Lagrangian. As seen in the previously-mentioned lattice works, the dynamics of the chiral phase transition in particular are sensitive to the masses of the lightest quark flavors. A full treatment of a theory with nonzero quark masses would require the following modifications to our dyon partition function: \\
(i) A generalization of the perturbative quark potential (\ref{eq_pot}) to arbitrary quark mass, the form of which is given in Ref. \cite{Fukushima:2017csk}. \\
(ii) A generalization of the hopping matrix elements $T_{ij}$ to arbitrary quark mass, which is not yet known. \\
(iii) The inclusion of quark mass terms on the diagonal elements of the hopping matrix as shown in Eq. (\ref{eq_mat}).

\begin{figure*}
	\includegraphics[width=0.95\linewidth]{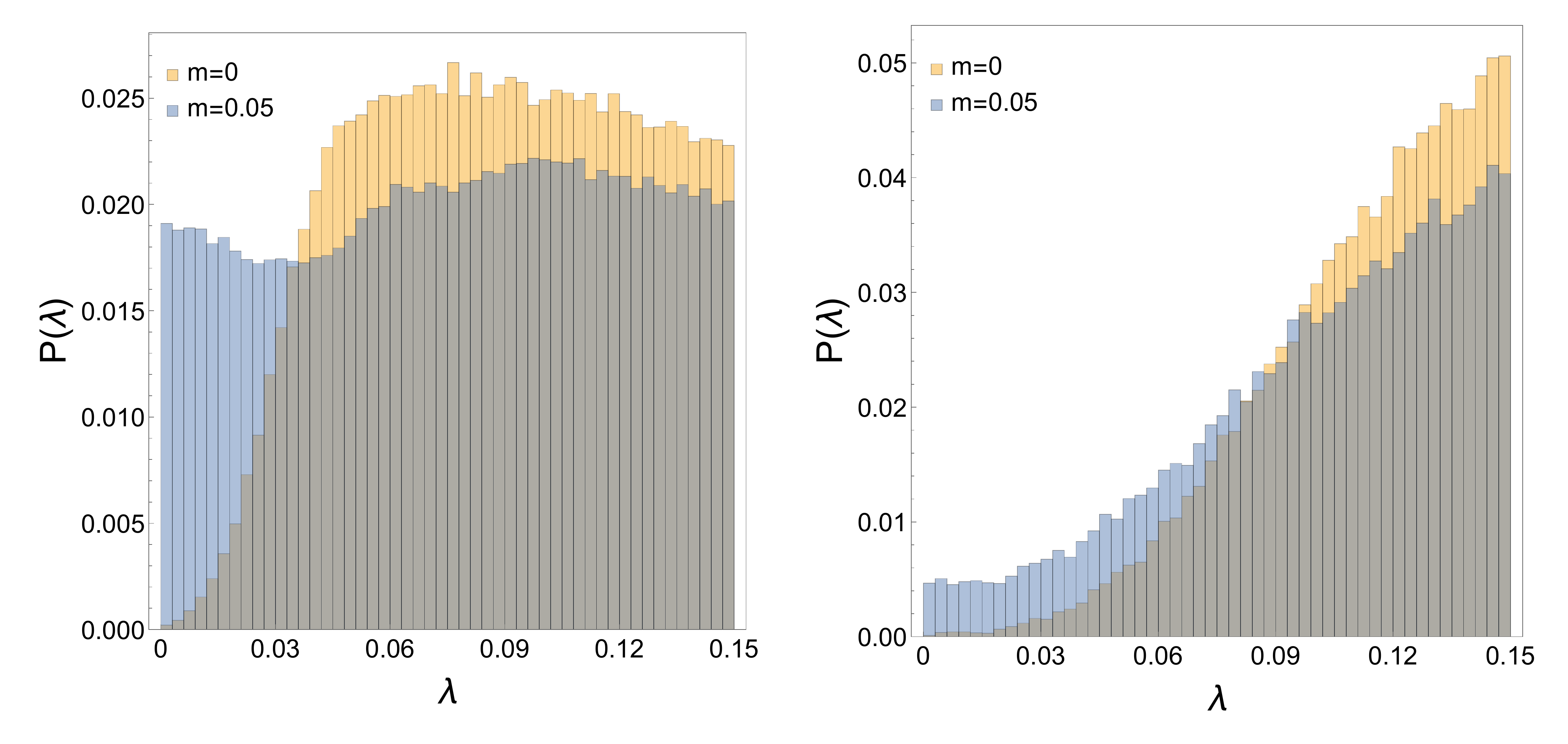}
	\caption{(Color online) Normalized probability distributions of eigenvalues in the near-zero-mode zone for both massless and $m=0.05$ quarks. Left: $S_0=8$, Right: $S_0=14$.}
	\label{fig_hist2} 
\end{figure*}

We do not do such a full treatment in this work. Instead we include only the quark mass term on the diagonals of the hopping matrix to demonstrate the qualitative impact it has on the eigenvalue spectrum. The nonzero quark mass effectively adds new diagrams to the fermionic determinant in which single dyons are allowed to have closed loops with a mass insertion that flips the chirality of the quark. The mass mediates the behavior of the quark-induced potential, driving it to a constant value at large distances, rather than remaining linear as in the massless case. 

Eigenvalue spectra for a small, nonzero quark mass are compared to the massless case in Fig. \ref{fig_hist2}. The nonzero quark mass allows for near-zero eigenvalues even at finite volume. The quark mass smooths the distribution of eigenvalues in the vicinity $\lambda \sim m$. At eigenvalues $\lambda > m$, the distributions are the same as in the massless case. In the broken phase, increasing the quark mass reduces the value of the condensate. It is known for example that the strange quark condensate is smaller than the up quark condensate \cite{Dominguez:2007hc}. The increased mass causes the condensate to decrease more slowly, becoming an analytic crossover with increased (now psuedocritical) $T_c$. One can see from Fig. \ref{fig_hist2} (right) that there is a nonzero condensate in the nonzero mass case, while it has already reduced to zero in the massless case. The finite-mass condensate never \textit{exactly} goes to zero, as there is no longer an exact symmetry. 

\section{Summary and Discussion}

In this work we have performed numerical simulations of a semiclassical ensemble of $SU(3)$ instanton dyons with $N_f=2$ flavors of massless quarks. Integration over the dyon degrees of freedom was performed in a periodic 3D box via Monte-Carlo methods. From tens of thousands of simulations, the properties of the ensemble in the thermodynamic limit were determined. 

The main addition to the simulations stemming from the inclusion of quarks is the computation of the fermionic determinant. This is approximated by the so-called hopping matrix, which contains only the subspace spanned by the quark zero modes. Quarks then 'hop' between $L$ and $\bar{L}$ dyons generating an effective interaction between all such dyons in the ensemble. These interactions, which can be dominated by single $L\bar{L}$ pairs or collective modes involving the overlap of many zero modes (see Fig. \ref{fig_loop}), determine the state of the chiral symmetry breaking.

The two phase transitions -- deconfinement and chiral symmetry restoration -- were observed. Confinement is studied by the value of the average Polyakov loop $\langle P(T) \rangle$, seen in Fig. \ref{fig_poly}. Indeed we find that the inclusion of quarks changes the deconfinement transition from first order to one compatible with that of the second-order transition in the $O(4)$ universality class. 

The near-zero-mode zone of the Dirac eigenvalue spectrum (Fig. \ref{fig_hist}) is used to determine the zero-eigenvalue density. This is directly related to the quark condensate via the Banks-Casher relation (\ref{eq_bcr}). We performed simulations at three different ensemble sizes in order to observe the non-trivial volume dependence of the spectra (Fig. \ref{fig_gaps}) and measure both the quark condensate and the eigenvalue gap as functions of the temperature. Fig. \ref{fig_cond} shows that both observables see nearly-simultaneous transitions to/from zero giving consistent determinations of the critical temperature $T_c$. Finally, we show that a small, but nonzero quark mass smooths the distribution and produces near-zero eigenvalues at higher temperatures, increasing $T_c$. 

Our results suggest that both phase transitions are driven primarily by the dyon densities. Confinement requires a sufficient density of dyons to overcome the perturbative quark and gluon interactions and shift the minimum to the confining holonomy. Chiral symmetry is similarly broken by a large density of $L$ and $\bar{L}$ dyons causing significant overlap between zero modes, leading to a dominance of large quark hopping loops, producing near-zero eigenvalues. 

Let us conclude with some discussion of the dyon model itself. It should be reminded that, as with the pure $SU(3)$ theory, our model contains parameters related to the short-range classical interactions (namely $V_0$ and $x_0$) which are phenomenological choices not known from first principles. With the hopping matrix elements we choose a simple parameterization that does not include the effects of interference from nearby $M_i$ dyons. Both of these aspects of the model should be studied more rigorously in order to improve the quantitative predictions of the model.

While we have been able to identify both phase transition, compared with the pure $SU(3)$ work \cite{DeMartini:2021dfi} the lack of a jump in the order parameters makes precise determinations of the critical temperatures more difficult. The deconfinement transition is slow and fits to potential forms yield a large variance in $T_{deconf}$. With the chiral transition, the drop in the condensate is much sharper, but requires better control of finite volume effects and different methods of extrapolation can modestly modify the determination of $T_c$ (see Appendix B). 

Despite its relative simplicity, there are some advantages to the dyon model compared with analogous lattice studies. The most obvious is the computational cost. Our largest simulations with $N_D \simeq 360$ involve integration over $\sim 1000$ degrees of freedom, while modern lattice simulations can involve some hundreds of millions of degrees of freedom. Additionally our largest simulations contain $\mathcal{O}(100)$ instantons at a time, significantly more than the number of instantons that are observed on a single time slice of lattice simulations. Thus, our simulations correspond to much larger spatial volumes than are used on the lattice. Lastly we are able to work directly with massless quarks, where the recent lattice results \cite{Ding:2018auz,HotQCD:2019xnw,Clarke:2019tzf,Ding:2020eql,Kaczmarek:2020sif,Clarke:2020htu} require nonzero quark masses and an extrapolation to the chiral limit.  

   \begin{acknowledgments}
	This work is supported by the Office of Science, U.S. Department of Energy under Contract No. DE-FG-88ER40388. The authors also thank the Stony Brook Institute for Advanced Computational Science for providing computer time on the SeaWulf computing cluster. 
   \end{acknowledgments}
   
    \appendix
    \section{Quark Zero Mode Density}
    
    Here we present a discussion of the general form for the quark zero mode density following the work in Ref. \cite{Chernodub:1999wg}. A detailed derivation of the gauge field and zero mode solutions are quite involved, requiring a combination of the Nahm transformation \cite{Nahm:1979yw} and ADHM construction \cite{Atiyah:1978ri}. We will simply present here the results that are most relevant to this work; we will give the explicit forms for the $SU(3)$ gauge group and antiperiodic quarks. 
    
    We remind again that the Polyakov loop at infinity has the holonomy phases (eigenvalues)
    \begin{equation}
    \mu_1 \le \mu_2 \le \mu_3 \le \mu_4 = \mu_1 + 1,
    \end{equation}
     and the dyon holonomies are defined as $\nu_i = \mu_{i+1} - \mu_i$. In terms of the single holonomy parameter in this work, the phases are $\mu_1 = -\nu$, $\mu_2 = 0$, $\mu_3 = \nu$.
     
     The action density of the gauge field can be written in the simple form in terms of the positions of the constituent dyons $\vec{y}_i$,
     \begin{equation}
     \begin{aligned}
     &tr F_{\mu\nu}^2 = \partial_{\mu}^2 \partial_{\nu}^2 \ln \psi, \\
     &\psi = \frac{1}{2} tr(\mathcal{A}_3 \mathcal{A}_2 \mathcal{A}_1) - \cos (2 \pi \tau T), \\
     & \mathcal{A}_i = \frac{1}{r_i} \begin{pmatrix}
     r_i & |\vec{y}_i - \vec{y}_{i+1}| \\
     0 & r_{i+1} \\
     \end{pmatrix} \begin{pmatrix}
     \cosh (R_i) & \sinh (R_i) \\
     \sinh (R_i) & \cosh (R_i) \\
     \end{pmatrix},
     \end{aligned}
     \end{equation}
     where $r_i = |\vec{x} - \vec{y}_i|$ and $R_i = 2 \pi \nu_i r_i$. The index $i$ is cyclical, e.g. $r_4 = r_1$. Plotting the action density numerically, one can see the interference effects of nearby dyons and in particular, that the dependence on Euclidean time $\tau$ vanishes when the dyons are all well separated.  
     
     The quark zero mode density can be written in a remarkably similar framework, in terms of the same matrices $\mathcal{A}_i$. Quarks with the boundary condition $\Psi_0(\vec{x},\tau)= e^{2 \pi i z} \Psi_0(\vec{x},\tau+\beta)$ have a zero-mode density
     \begin{equation}
     |\Psi_0(x)|^2= \frac{-1}{4\pi^2} \partial_{\mu}^2 \hat{f}(x).
     \end{equation}
     For the antiperiodic quarks, $z=1/2$ and $\hat{f}(x)$ is
     \begin{equation}
     \hat{f}(x)= \frac{\pi}{r_3 \psi} \langle v_3|\mathcal{A}_2 \mathcal{A}_1 \mathcal{A}_3|w_3 \rangle,
     \end{equation}
     where $v_3$ and $w_3$ are the 2-component spinors with elements
     \begin{equation}
     \begin{aligned}
     v_3^1 &= -w_3^2 = \sinh (2\pi(\frac{1}{2}-\nu)r_3), \\
     v_3^2 &= w_3^1 = \cosh (2 \pi (\frac{1}{2} - \nu) r_3).
     \end{aligned}
     \end{equation}
     The quark zero mode is localized on the dyon $i$ such that $\mu_i < z < \mu_{i+1}$, which is the $L$ dyon associated with $i=3$. 
     
     Taking the necessary derivatives to write out either density directly in terms of the dyons' coordinates, even after simplification, results in formulae which are dozens of lines long as Mathematica outputs. These are far too complicated to compute at each step of a Metropolis update, hence the use of a simple parameterization in this work. In the limit that the dyons are well separated, one can see that the general zero-mode density given here is exactly what is predicted by Eq. (\ref{eq_zero}), up to an overall normalization constant. 
     
\section{Comparison of Infinite-Volume Extrapolations} \label{app_extra}         
     
The determination of the chiral condensate depends a choice of interpolation function between different system sizes (\ref{eq_int}). Rather than interpolating between all three volumes, one could consider using just two of the volumes. In particular we consider scaling directly between volumes $V$ and $3V$ and scaling between $2V$ and $3V$ with the functions
\begin{equation}
	\Sigma(V\rightarrow3V) = \Sigma_2 \left( \frac{\Sigma_1^{3V}}{2\Sigma_1^{V}} - \frac{1}{2} \right),
\end{equation}  
   \begin{equation}
   	\Sigma(2V\rightarrow3V) = \Sigma_2 \left(\frac{2\Sigma_1^{3V}}{\Sigma_1^{2V}} - 2 \right).
   \end{equation}  
As with the interpolating function used in the main text, we set a maximum value of 1 on the scaling factors.

\begin{figure}[h]
	\includegraphics[width=0.85\linewidth]{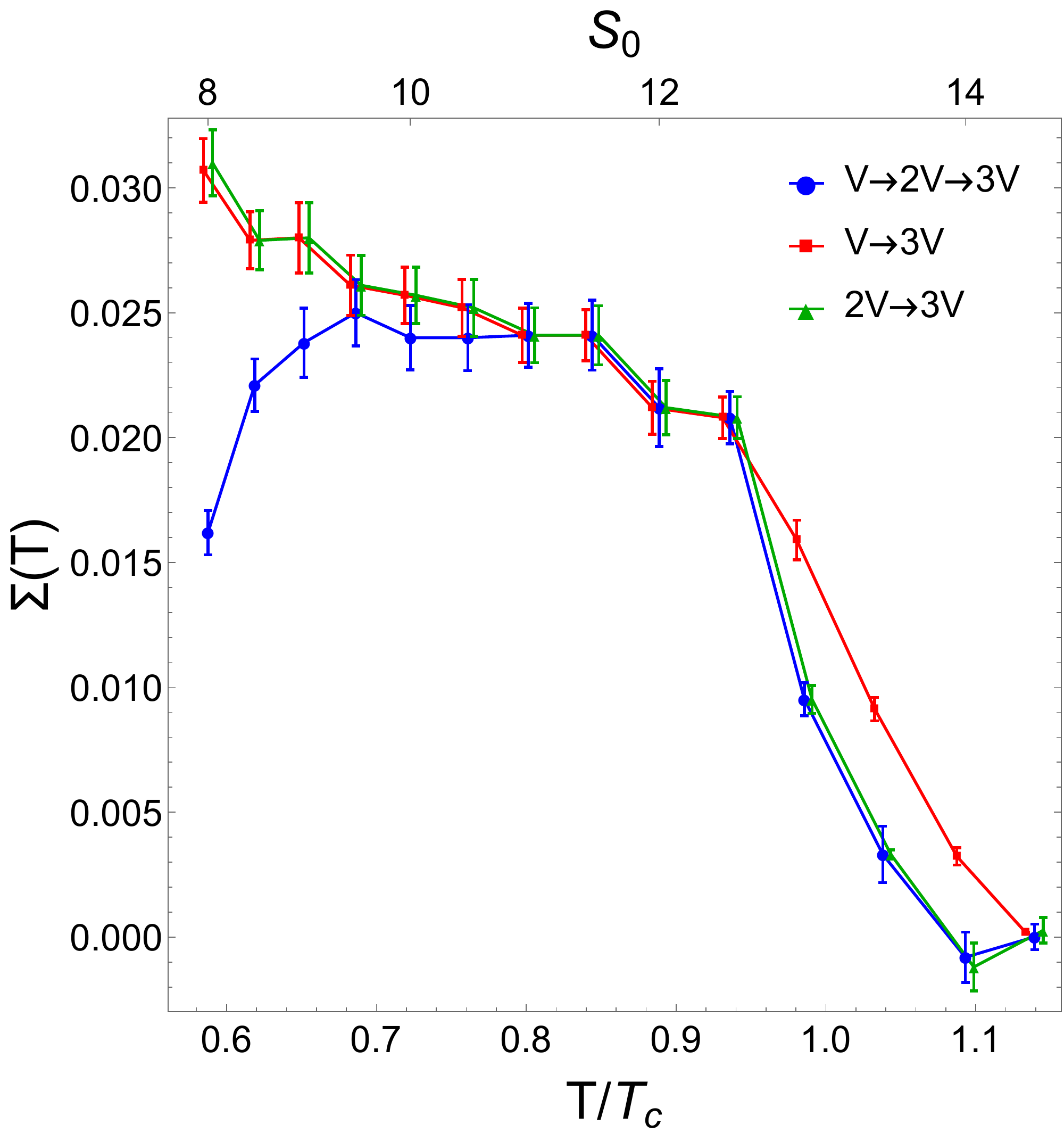}
	\caption{(Color online) The chiral quark condensate $\Sigma(T)$ determined from three different interpolating functions. Points shifted slightly horizontally for readability.} 
	\label{fig_comp}
\end{figure}

Each of these functions results in different results for the chiral condensate and are compared in Fig. \ref{fig_comp}. There are a few qualitative differences between them. For several of the data points, all three give the exact same answers as the range of the suppressed region scales faster than $1/V$ and the scaling functions are all taking on the ceiling value of 1. At the lowest values the function using all three volumes decreases slowly rather than increasing like the other two. This is because the scaling factor from $V$ to $2V$ is less than 1 while the others are 1. This may be an indication that at these temperatures, where the density is highest and the volumes are smallest, the system may not be large enough for the mesoscopic scaling relations (\ref{eq_rmt}) to apply. One should note that it is not the total number of dyons $N_D$, but the number of zero modes $2N_L$ which is relevant to the eigenvalue distributions. For the smallest ensemble sizes, $2N_L \sim 20-25$.  

From Fig. \ref{fig_comp} one can see that the value of $T_c$ can vary by about $10\%$, depending on the choice of interpolating function. Of course, one could also consider other functions which may produce even more varied results. 

One could also consider using only two ensemble sizes for the linear fit to the eigenvalue gap data. Doing so however, one finds much less dependence on the choice of which sizes to include. At low $T$ the gaps also show better agreement with the expected $1/V$ scaling and do not need an interpolating function. Thus we conclude that the eigenvalue gap $\Delta(T)$ is a more stable and reliable indicator of which chiral phase the system is in. Of course, the best way to improve the results for either observable is to continue to go to larger volumes. 
     
\bibliography{dyons_w_quarks}

\end{document}